\documentclass[aps,prx,reprint,twocolumn, amsmath,amssymb,amsfonts,footinbib,
superscriptaddress,longbibliography,
notitlepage]{revtex4-1}

\AtBeginDocument{\usepackage{booktabs}}               
\makeatletter
\g@addto@macro\bfseries{\boldmath}
\makeatother

\usepackage[varg]{txfonts}
\usepackage[T1]{fontenc}
\usepackage[utf8]{inputenc}

\usepackage[varg]{txfonts}
\usepackage{hyperref}
\usepackage{amsmath}
\usepackage{color}
\usepackage{graphicx}
\usepackage[percent]{overpic}
\usepackage{mathrsfs}
\usepackage{bm}
\usepackage{braket}
\usepackage{mathtools}

\definecolor{cred}{RGB}{228,26,28}
\definecolor{cblue}{RGB}{8,48,107}
\definecolor{cgreen}{RGB}{77,175,74}
\definecolor{cgray}{RGB}{150,150,150}
\definecolor{clgray}{RGB}{200,200,200}
\definecolor{cpurple}{RGB}{152,78,163}
\definecolor{corange}{RGB}{255,127,0}
\definecolor{cgold}{RGB}{230,171,2}

\newcommand{\avg}[1]{\langle #1 \rangle}

\newcommand{\h}[1]{{#1}^{\dagger}}

\newcommand{\del} {\partial}

\newcommand{\re}{{\rm Re}}
\newcommand{\im}{{\rm Im}}

\newcommand{\subref}[2]{\ref{#1}\hyperref[#1]{#2}}

\renewcommand{\vec}[1]{\boldsymbol{#1}}
\renewcommand{\l}[1]{{#1}^{\phantom{\dagger}}}

\newcommand{\vhat}[1]{\vec{\hat{#1}}}

\newcommand{\curv}{C}

\definecolor{cred}{RGB}{188,55,84}
\hypersetup{colorlinks=true,linkcolor=cred,citecolor=cred,urlcolor=cred}

\begin{document}

\title{Pseudo-Goldstone modes and dynamical gap generation from order-by-thermal-disorder}
\author{Subhankar Khatua}
%\email{skhatua1@uwaterloo.ca}
\affiliation{Department of Physics, University of Windsor, 401 Sunset Avenue, Windsor, Ontario, N9B 3P4, Canada}
\affiliation{Department of Physics and Astronomy, University of Waterloo, Waterloo, Ontario, N2L 3G1, Canada}
\author{Michel J. P. Gingras}
%\email{gingras@uwaterloo.ca}
\affiliation{Department of Physics and Astronomy, University of Waterloo, Waterloo, Ontario, N2L 3G1, Canada}
\author{Jeffrey G. Rau}
%\email{jrau@uwindsor.ca}
\affiliation{Department of Physics, University of Windsor, 401 Sunset Avenue, Windsor, Ontario, N9B 3P4, Canada}

\begin{abstract}
Accidental ground state degeneracies -- those not a consequence of global symmetries of the Hamiltonian -- are inevitably lifted by fluctuations, often leading to long-range order, a phenomenon known as ``order-by-disorder’’ (ObD). 
The detection and characterization of ObD in real materials currently lacks clear, qualitative signatures that distinguish ObD from conventional energetic selection. 
We show that for order-by-\emph{thermal}-disorder (ObTD) such a signature exists: a characteristic temperature dependence of the fluctuation-induced pseudo-Goldstone gap.
We demonstrate this in a minimal two-dimensional model that exhibits ObTD, 
the ferromagnetic Heisenberg-compass model on a square lattice.
Using spin-dynamics simulations and self-consistent mean-field calculations, we determine the pseudo-Goldstone gap, $\Delta$, and show that at low temperatures it scales as the square root of temperature, $\sqrt{T}$.
We establish that a power-law temperature dependence of the gap is a general consequence of ObTD, showing that all key features of this physics can be captured in a simple model of a particle moving in an effective potential generated by the fluctuation-induced free energy.
\end{abstract}

\date{\today}

\maketitle

Strongly competing interactions, or frustration, enhance quantum and thermal fluctuations, and undermine the development of conventional magnetic order. 
The latter can even be prevented entirely down to zero temperature, leading to classical~\cite{Anderson_spinels,Villain_ZPhysB,Moessner-Chalker} or quantum spin liquids~\cite{Canals,Kitaev,Gingras_McClarty,Savary_Balents,Imai-Lee,Knolle_Moessner,Balents2010}. 
However, additional perturbative interactions can relieve the frustration and favor the development of long-range order (LRO). 
Accordingly, the majority of spin liquid candidates ultimately evade fate as a spin liquid~\cite{Imai-Lee,Wen2019}. 
The ability of such perturbative interactions, largely inconsequential without frustration, to dictate the ground state and low-temperature properties of a system is at the root of the plethora of exotic phenomena displayed by highly-frustrated magnetic materials~\cite{Balents2010,Lacroix2011,Springer-spin-ice,Gardner2010,Hallas-AnnRevCMP,Rau-AnnRevCMP,Trebst2022,Takagi2019}.

This relief of frustration is not always complete. 
Instead of an extensively degenerate manifold, a system can possess a sub-extensive \emph{accidental} ground state degeneracy, unprotected by symmetry. 
Classically, this degeneracy can be robust to a range of realistic interactions including symmetry-allowed two-spin exchange~\cite{Savary2012}. 
Here, the role of fluctuations is dramatically changed: instead of being detrimental, they can lift the classical degeneracy and stabilize order -- this is the celebrated phenomenon of \emph{order-by-disorder} (ObD)~\cite{Villain1980,Shender1982,Henley1989}. 
While numerous theoretical models have been proposed~\cite{tessman1954,Villain1980,Shender1982,Henley1989,Prakash1990,Kubo1991,Chubukov1992,Reimers1993,Reimers1994,Henley1994,Champion2003,Baskaran2008,McClarty2014,Danu2016}, there is a paucity of real materials that unambiguously harbor ObD~\cite{Brueckel1988,Kim1999,Savary2012,Ross2014,Sarkis}.
The standard strategy for experimental confirmation of ObD is indirect, relying on parametrizing a theoretical model of the material, establishing ObD within that model, and then validating its predictions for the ordered state experimentally.

While this program has been applied somewhat successfully to a handful of materials~\cite{Brueckel1988,Kim1999,Savary2012,Ross2014,Sarkis}, the inability to evince ObD directly, without relying on detailed modelling, highlights something lacking in our understanding of ObD. 
Clear, \emph{qualitative}, and model-independent signatures are needed; for example, experimental observation of characteristic power-laws in heat capacity or transport can diagnose the character of low-energy excitations, 
such as exchange statistics, dimensionality or their dispersion relations~\cite{Ashcroft-Mermin,Knolle_Moessner,Wen2019,X-G-WenBook}. 
Does the presence of ObD exhibit a ``smoking-gun'' experimental signature? This can be difficult or subtle to discern. For ObD from quantum fluctuations~\cite{Shender1982}, the formation of an ObD spin-wave gap is generally not distinguishable from one induced energetically by multi-spin interactions~\cite{McClarty_2009,Petit2014,Rau-Petit}.

%%%%%%%%%%%%% ---------------  %%%%%%%%%%%%%

\begin{figure*}
        \centering
        \includegraphics[width = \textwidth]{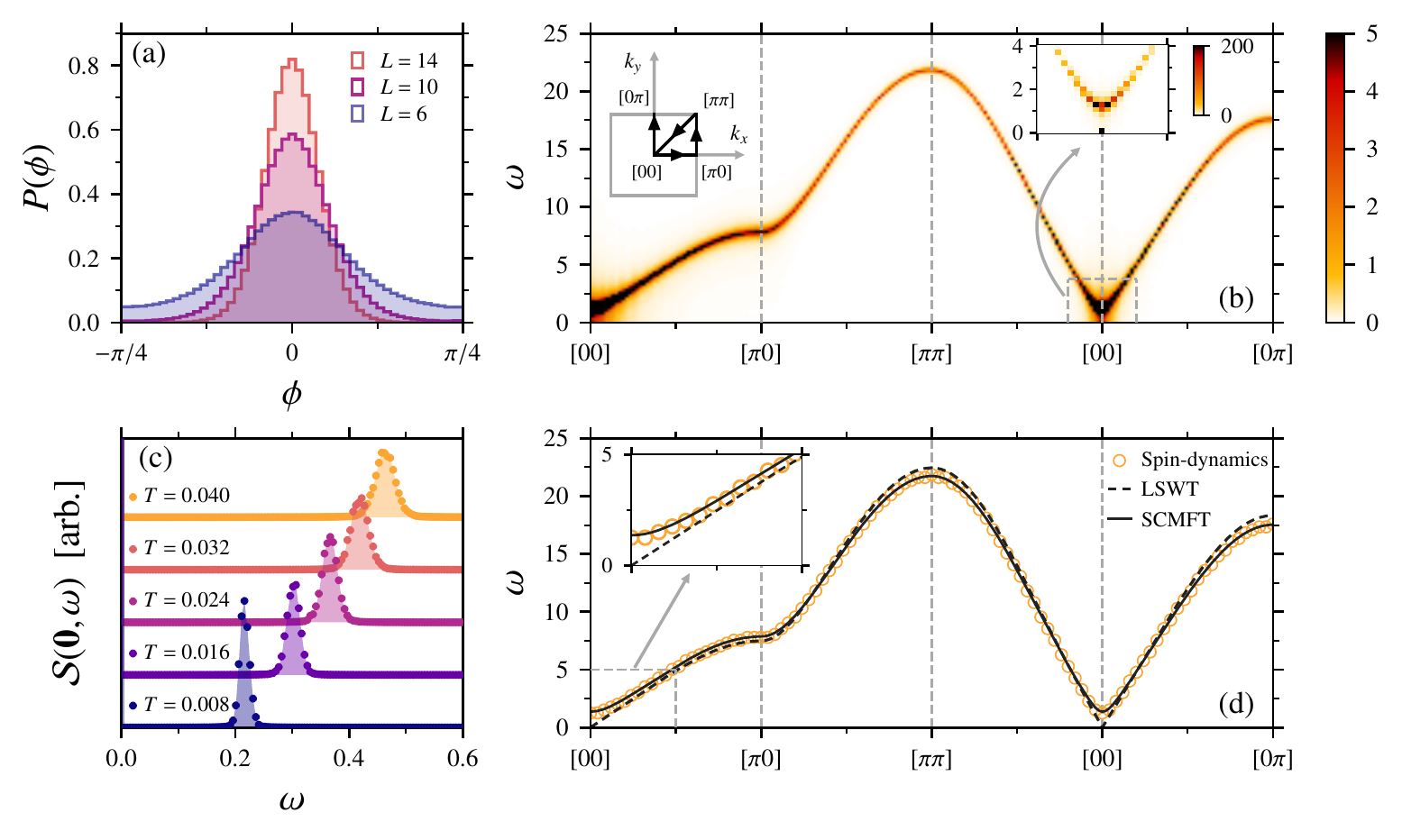}
        \caption[]{(a) Probability distribution, $P(\phi)$, of the angle, $\phi$, characterizing the direction of the net magnetization obtained using MC simulations with $K = 5$ at $T = 0.4$ for several system sizes, $L$. Due to $C_4$ symmetry, $P(\phi)$ is shown for $\phi\in[-\pi/4,\pi/4]$. (b) Dynamical structure factor, ${\cal S}(\vec{k},\omega)$ obtained from spin-dynamics simulations for $L=100$ with $K = 5$ at $T = 0.4$ along a path through the Brillouin zone (see left inset). The overall intensity is arbitrary. (Right inset) Spectrum near $[00]$ showing the PG gap~\cite{notefig1}. (c) Dynamical structure factor at $\vec{k} = \vec{0}$, ${\cal S}(\vec{0},\omega)$, obtained from spin-dynamics simulations for $L=40$ at various temperatures with $K = 5$. As above, the overall intensity is arbitrary. (d) Excitation spectrum along the same path as in panel-(b) from the LSWT, SCMFT, and spin-dynamics simulations with $K = 5$ for $L=100$ at $T = 0.4$. The spin-dynamics spectrum tracks the frequencies of maximum of ${\cal S}(\vec{k},\omega)$. The inset highlights a small region near $[00]$, showing the PG mode. } 
        \label{fig.spectrum_mod}
    \end{figure*}

In this Letter, we identify a clear signature of order-by-\emph{thermal}-disorder (ObTD): a dynamically generated gap growing as the square root of temperature. 
We investigate this gapped ``pseudo-Goldstone'' (PG) mode~\cite{Weinberg1972,Burgess2000,Nitta2015} in a minimal 2D classical spin model exhibiting ObTD, the ferromagnetic Heisenberg-compass model on a square lattice, belonging to a class of models relevant to Mott insulators with strong spin-orbit coupling~\cite{Nussinov2015,Dorier2005,Jackeli2009,Trousselet2010,Trousselet2012,Boseggia2013,Katukuri2014,Vladimirov2015,Zhang2016}.
Through spin-dynamics simulations, we determine the PG gap, $\Delta$, and show it varies with temperature as $\cramped{\Delta \propto \sqrt{T}}$, in quantitative agreement with the self-consistent mean-field theory (SCMFT) that we present. 
This mode is well-defined, with its linewidth, $\Gamma$, due to thermal broadening, $\cramped{\Gamma \propto T^2 \ll \Delta}$. 
We further demonstrate that our key results can be captured by an effective description of a particle moving in a potential generated by the fluctuation-induced free energy.
Using this picture, we argue that the temperature dependence of the PG gap, $\sqrt{T}$ ($T$) for type-I (II) PG modes,
which have dispersion 
$\cramped{\omega \propto |\vec{k}|}$
and 
$\cramped{\omega \propto |\vec{k}|^2}$,
respectively~\cite{watanabe2020counting}, is \emph{universal} and applicable to \emph{any} system exhibiting ObTD.
Finally, due to the low dimensionality~\cite{Coleman1973} of the 2D model considered, ObTD faces a subtle competition against potentially infrared-divergent fluctuations~\cite{Mermin1966,Hohenberg1967}. While ObTD ultimately prevails and true LRO develops, the magnetization displays logarithmic corrections at low temperatures, a remnant of the diverging infrared fluctuations.

\textit{Model.---} We consider the \emph{classical} ferromagnetic Heisenberg-compass model on a square lattice 
   \begin{equation}
   \label{eq.Ham}
  {\cal H} = \sum_{\vec{r}}\biggl[-J\sum_{\vec{\delta} = \vec{x},\vec{y}} 
    \vec{S}^{\phantom{x}}_{\vec{r}\phantom{\vec{}}} \cdot \vec{S}^{\phantom{x}}_{\vec{r}+\vec{\delta}\phantom{\vec{}}}
    -K \left(S^x_{\vec{r}\phantom{\vec{}}} S^x_{\vec{r}+\vec{x}} + S^y_{\vec{r}\phantom{\vec{}}} S^y_{\vec{r}+\vec{y}} 
  \right)\biggr],
\end{equation}
where $\vec{S}_{\vec{r}}\equiv (S^x_{\vec{r}}, S^y_{\vec{r}},S^z_{\vec{r}})$ is a unit vector at site $\vec{r}$, and $\vec{\delta} = \vec{x},\vec{y}$ 
denote the nearest-neighbor bonds. We consider ferromagnetic Heisenberg and compass interactions with  $J>0$, $K>0$ (see SM~\cite{SM} for a discussion of other signs) and with $J$ the unit of energy, setting $J \equiv \hbar \equiv k_{\rm B} \equiv 1$ throughout.  

For $K=0$, the model~[Eq.\eqref{eq.Ham}] is the well-known Heisenberg ferromagnet with uniform ferromagnetic ground states of arbitrary direction, $\cramped{\vec{S}_{\vec{r}}  = \vhat{n}}$, related by global spin-rotation symmetry.
For $K > 0$, this symmetry is absent and ${\cal H}$ in Eq.~\eqref{eq.Ham} is minimized by any uniform magnetization in the $\vhat{x}-\vhat{y}$ plane. These ground states are characterized by an angle $\phi \in [0,2\pi)$ with $\vec{S}_{\vec{r}} = \cos{\phi}\,\vhat{x}+\sin{\phi}\,\vhat{y}$. 
Unlike the pure Heisenberg ferromagnet, these are only \emph{accidentally} degenerate, as continuous in-plane spin rotations do not preserve the anisotropic compass term. 
However, a discrete $C_4$ symmetry about the $\vhat{z}$ axis and $C_2$ symmetries about the $\vhat{x}$ and $\vhat{y}$ axes still remain.   

\textit{Simulations.---} We first demonstrate that this model exhibits ObTD via Monte Carlo (MC) simulations on a lattice with $N = L^2$ sites.
To expose the state selection, we construct a probability distribution for the magnetization orientation $\phi$, $P(\phi)$, using a sample of thermalized states (see SM~\cite{SM}). As shown in Fig.~\subref{fig.spectrum_mod}{(a)}, $P(\phi)$ exhibits maxima at $\phi = 0$, $\pi /2$, $\pi$, $3\pi/2$, corresponding to ferromagnetic ground states with $\vhat{n}$ along the $\pm \vhat{x}$, $\pm\vhat{y}$ directions. At low temperatures, fluctuations thus select four discrete ground states via ObTD from a one-parameter manifold of states.

We now consider the classical dynamics to examine the associated PG mode. The equation of motion for the classical spins is the Landau-Lifshitz equation~\cite{Landau1935}, $d\vec{S}_{\vec{r}}/dt~=~\vec{B}_{\vec{r}} \times \vec{S}_{\vec{r}}$, describing precession about the exchange field, $\vec{B_r}$, produced by neighboring spins
\begin{equation}
\vec{B}_{\vec{r}} \equiv -\!\!\sum_{\vec{\delta}=\pm \vec{x},\pm \vec{y}}  \left[J\vec{S}^{\phantom{\delta}}_{\vec{r}+\vec{\delta}}+K S^{\delta}_{\vec{r}+\vec{\delta}} \vec{\delta} \right].
\end{equation}
Starting with states drawn via MC sampling at temperature $T$, we numerically integrate the Landau-Lifshitz equations, and compute the dynamical structure factor, ${\cal S}(\vec{k},\omega) = \avg{|\vec{S}_{\vec{k}}(\omega)|^2}$, where $\vec{S}_{\vec{k}}(\omega)$ is the Fourier transform of the spins, and $\avg{\cdots}$ denotes averaging over the initial states~\cite{SM}. 
Results for ${\cal S}(\vec{k},\omega)$ at a representative $T$ and $K$~\cite{SM} are shown in Fig.~\subref{fig.spectrum_mod}{(b)}, exhibiting sharp spin-waves with a \emph{nearly} gapless mode at $\vec{k}=\vec{0}$. Closer examination reveals a well-defined gap, as highlighted in the top right inset of Fig.~\subref{fig.spectrum_mod}{(b)} -- this is the PG gap. 

\begin{figure}
\includegraphics[width=\columnwidth]{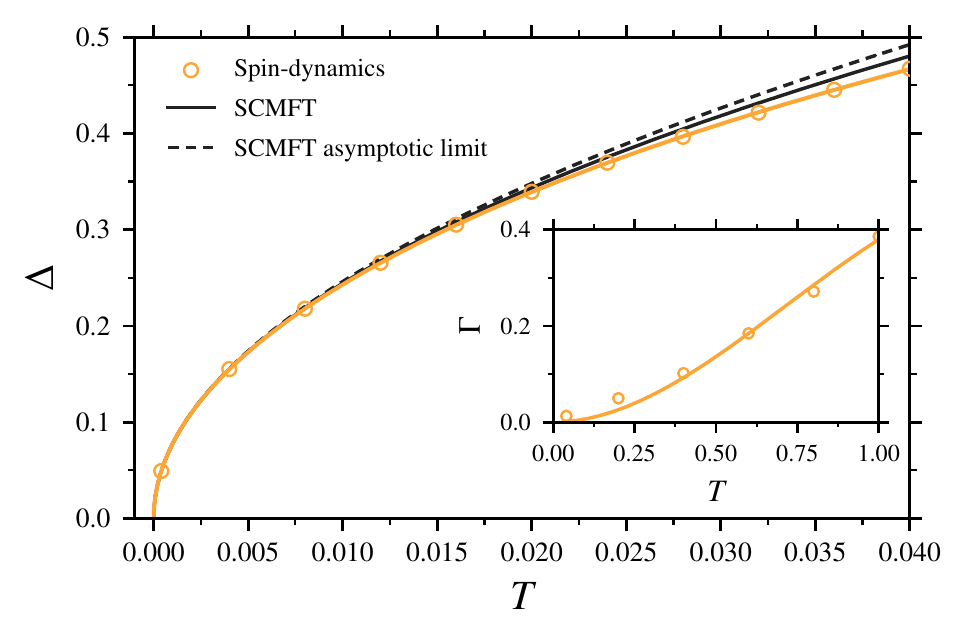}
\caption{Pseudo-Goldstone gap, $\Delta$, as a function of temperature from spin-dynamics simulations with $K=5$. The data is well-described by the fit $\Delta = 2.46242\sqrt{T}- 3.21907\, T^{3/2}$.
The SCMFT gap agrees with it quantitatively and provides the asymptotic $T \rightarrow 0$ scaling, $2.46147 \sqrt{T}$. The asymptotic limit of the SCMFT gap is obtained from a calculation at a fixed low temperature $T = 0.001$. (Inset) Linewidth of the PG mode, $\Gamma$, as a function of temperature from spin-dynamics simulations. It is well described by the fit, $\Gamma = 0.709286\,T^2 - 0.329751\,T^3 
$. All data have been extrapolated in the system size to the thermodynamic limit~\cite{SM}. 
}
\label{fig.gap}
\end{figure}

To determine the PG gap \emph{quantitatively}, we consider a cut of the structure factor at $\vec{k}=\vec{0}$, i.e., ${\cal S}(\vec{0},\omega)$.
As the PG gap is much smaller than the bandwidth of the spectrum [see Fig.~\subref{fig.spectrum_mod}{(b)}], a significantly higher frequency resolution is required to accurately compute the gap~\cite{SM}, so a much longer integration time window is necessary.
Cuts, ${\cal S}(\vec{0},\omega)$, for several temperatures are presented in Fig.~\subref{fig.spectrum_mod}{(c)}, with the peak location indicating the PG gap (see SM~\cite{SM}). The temperature dependence of $\Delta$ is shown in Fig.~\ref{fig.gap}.
The leading contribution to the PG gap scales as the \emph{square root} of temperature, vanishing as $T\rightarrow 0$, being well-described by the fit $\Delta \sim 2.46\sqrt{T}$.

The thermal broadening of the spectrum induces a finite width to all excitations, including the PG mode. 
The PG mode linewidth, $\Gamma$, can be obtained from the full-width at half maximum of ${\cal S}(\vec{0},\omega)$ [see Fig.~\subref{fig.spectrum_mod}{(c)}] as a function of temperature. 
The inset in Fig.~\ref{fig.gap} shows that $\Gamma \propto T^2$ at low temperatures (see SM~\cite{SM}). Since $\Gamma \ll \Delta$ as $T \rightarrow 0$, this PG mode is well-defined.

\textit{Spin-wave analysis.---} The simulations reveal that the system displays LRO and hosts a PG excitation, where the PG gap and linewidth scale with temperature as $\sqrt{T}$ and $T^2$, respectively.
To understand how these scaling laws arise, we consider a spin-wave analysis about the ordered state~\cite{Auerbach1998}.
Since tackling spin-wave interactions is difficult within a purely classical approach~\cite{Martin1973,Deker1975,Enz1976}, we follow the more widely used and computationally convenient quantum spin-wave analysis~\cite{Bruus2004,Blaizot1986,Mahan2000}, taking the classical limit only at the end.

We first discuss the spectrum and state selection due to ObTD in linear spin-wave theory (LSWT).
Expanding about a classical ground state (parametrized by $\phi$) using the Holstein-Primakoff (HP) transformation~\cite{Auerbach1998}, we obtain to $O(S)$ 
\begin{equation}
{\cal H}_2 = \sum_{\vec{k}}\left[ A^{\phantom{\dagger}}_{\vec{k}} \h{a}_{\vec{k}} \l{a}_{\vec{k}} 
    + \frac{1}{2!}\left(B^{\phantom{\dagger}}_{\vec{k}} \h{a}_{\vec{k}} \h{a}_{-\vec{k}} +{\rm H.c.}\right)\right],
    \label{eq.Ham_quad}
\end{equation}
where $\l{a}_{\vec{k}}$ denotes the bosonic annihilation operator at wave vector $\vec{k}$, and $A_{\vec{k}}$ and $B_{\vec{k}}$ depend on $\phi$, $J$, and $K$ (see SM~\cite{SM}). 
${\cal H}_2$ in Eq.~\eqref{eq.Ham_quad} can be diagonalized by a Bogoliubov transformation~\cite{Auerbach1998}, giving spin-wave energies $\omega_{\vec{k}} = \sqrt{A_{\vec{k}}^2 - B_{\vec{k}}^2}$. 
As the spectrum depends on the ground state angle $\phi$, fluctuations can lift the accidental classical degeneracy. 
To examine state selection due to ObTD, we search for the ground states for which the free energy is minimal. Starting with the quantum free energy $F_{\rm{Q}} = \frac{1}{2}\sum_{\vec{k}}\omega_{\vec{k}}+ T\sum_{\vec{k}}\ln\left(1 - e^{-\omega_{\vec{k}}/T}\right)$, the classical limit, $T\gg \omega_{\vec{k}}$  yields
$F_{\rm{Q}}\rightarrow F = T \sum_{\vec{k}} \ln{\omega_{\vec{k}}}$~\cite{Kardar2007}. 
This classical free energy has four minima at $\phi = 0,\, \pi/2,\,\pi,\,3\pi/2$ -- establishing selection by ObTD, in agreement with the MC results.  

Within LSWT, quantum and classical calculations give the same spectrum, $\omega_{\vec{k}}$~\cite{Henley1989}. 
This spectrum, calculated about $\phi =0$, exhibits a gapless mode at $\vec{k} = \vec{0}$ as shown in Fig.~\subref{fig.spectrum_mod}{(d)}. 
To obtain a PG gap, spin-wave interactions must be included, as we next discuss. 

\textit{Interacting spin waves.---} Performing the HP expansion to next order in $1/S$, the LSWT Hamiltonian [Eq.~\eqref{eq.Ham_quad}] is augmented by interaction terms. 
Three-boson interactions are absent due to a $C_2$ symmetry about the ordering direction, leaving only terms quartic in the bosons at $O(S^0)$ (see SM~\cite{SM}). 
To treat this interacting problem, we adopt a mean-field approach~\cite{Bruus2004,Blaizot1986}, decoupling the quartic terms into products of quadratic terms and thermal averages of two-boson operators. 
Following this procedure, the new effective quadratic Hamiltonian mirrors Eq.~\eqref{eq.Ham_quad}, but with $A_{\vec{k}}$ and $B_{\vec{k}}$ replaced with $(A_{\vec{k}} + \delta A_{\vec{k}})$ and $(B_{\vec{k}} + \delta B_{\vec{k}})$. These corrections are 
\begin{align}
    \delta A_{\vec{k}} &= \frac{1}{N}\sum_{\vec{q}} \left[V^{\phantom{\dagger}}_{\vec{k},\vec{q},\vec{0}} \avg{\h{a}_{\vec{q}}\l{a}_{\vec{q}}}
                         +\frac{1}{2}\left( D^{\phantom{\dagger}}_{\vec{q},-\vec{q},\vec{k}} \avg{\h{a}_{\vec{q}} \h{a}_{-\vec{q}}}+{\rm c.c.}
                         \right)
 \right],\nonumber
        \\
        \delta B_{\vec{k}} &= 
        \frac{1}{N}\sum_{\vec{q}} \left[D^{\phantom{\dagger}}_{\vec{k},-\vec{k},\vec{q}} \avg{\h{a}_{\vec{q}}\l{a}_{\vec{q}}}+
        \frac{1}{2}
        V_{\vec{q},-\vec{q},\vec{k}-\vec{q}} \avg{a_{\vec{q}} a_{-\vec{q}}}
        \right],
        \label{eq.delA_delB}
\end{align}
where $V_{\vec{k}_1,\vec{k}_2,\vec{k}_3}$ and $D_{\vec{k}_1,\vec{k}_2,\vec{k}_3}$ are the coefficients for the 2-2 and 3-1 magnon scattering terms at $O(S^0)$~\cite{SM}, and $\avg{\cdots}$ is a thermal average. When these averages are computed using LSWT [Eq.~\eqref{eq.Ham_quad}], the corrections  [Eq.~(\ref{eq.delA_delB})] reproduce the results obtained from leading order perturbation theory~\cite{Loly1971,Chubukov1994}. However, because of the gapless mode, these $\delta A_{\vec{k}}$ and $\delta B_{\vec{k}}$ individually diverge in the ($T\gg \omega_{\vec{k}}$) classical limit and perturbation theory breaks down~\cite{SM}.

To resolve these divergences, we compute the averages in Eq.~\eqref{eq.delA_delB} using SCMFT, obtaining a renormalized spectrum, $\Omega_{\vec{k}}$ (see SM~\cite{SM}). 
Explicitly, $\avg{\h{a}_{\vec{k}}\l{a}_{\vec{k}}}$ and $\avg{\l{a}_{\vec{k}} \l{a}_{-\vec{k}}}$ are, classically, computed self-consistently (until convergence) using Eq.~\eqref{eq.delA_delB} and
\begin{align}
    \avg{\h{a}_{\vec{k}} \l{a}_{\vec{k}}} &= \frac{T (A_{\vec{k}} + \delta A_{\vec{k}})}{\Omega^2_{\vec{k}}}, &
    \avg{\l{a}_{\vec{k}} \l{a}_{-\vec{k}}}  &= -\frac{T(B_{\vec{k}} + \delta B_{\vec{k}})}{\Omega^2_{\vec{k}}},
    \label{eq.therm_avg}
\end{align}
where $\Omega_{\vec{k}} = \sqrt{(A_{\vec{k}} + \delta A_{\vec{k}})^2 - (B_{\vec{k}} + \delta B_{\vec{k}})^2}$ and $\avg{\l{a}_{\vec{k}} \l{a}_{-\vec{k}}}=  \avg{\h{a}_{\vec{k}} \h{a}_{-\vec{k}}}$. 

The SCMFT spectrum $\Omega_{\vec{k}}$, plotted in Fig.~\subref{fig.spectrum_mod}{(d)}, exhibits a clear gap at $\vec{k} = \vec{0}$. The PG mode, gapless in LSWT, has now become gapped due to magnon-magnon interactions. 
Excellent agreement between the spectra from SCMFT and spin-dynamics simulations is observed across the full Brillouin zone [see Fig.~\subref{fig.spectrum_mod}{(d)}]. 
The temperature dependences of $\Delta$ from these two approaches in Fig.~\ref{fig.gap} agree quantitatively, with the same $\sqrt{T}$ scaling as $T\rightarrow 0$. This is a key result of this work, establishing a clear spectral \emph{signature} of ObTD.

While the SCMFT is successful in describing the excitation energies, it does not address thermal broadening, since $\delta A_{\vec{k}}$ and $\delta B_{\vec{k}}$ are real, giving an infinite magnon lifetime. 
To obtain a finite linewidth, perturbation theory must be carried out to higher order. 
We expect that $\delta A_{\vec{0}} \equiv \delta A_{\vec{k}=\vec{0}}$ and $\delta B_{\vec{0}}\equiv \delta B_{\vec{k}=\vec{0}}$, interpreted as contributions to the magnon self-energy~\cite{SM}, can be expanded in $T$ as $\delta A_{\vec{0}} = a_1T + a_2T^2 +\cdots $ and $\delta B_{\vec{0}} = b_1T + b_2T^2 +\cdots $. 
Since $|A_{\vec{0}}| = |B_{\vec{0}}|$, reflecting the gapless LSWT spectrum, and $a_1$, $b_1$ [the $O(T)$ corrections in Eq.~\eqref{eq.delA_delB}] are real; any imaginary part, and thus finite lifetime, must arise from $a_2$ or $b_2$. 
Expanding $\Omega_{\vec{0}}\equiv \Omega_{\vec{k}=\vec{0}}$ in $T$ yields  $\im\, \Omega_{\vec{0}} \approx (\im \,a_2)\, T^2 + \cdots$ (see SM~\cite{SM}).
The real part, $\re\,{\Omega_{\vec{0}}}$, maintains its leading $\sqrt{T}$ dependence (providing the PG gap) while $\im\,{\Omega_{\vec{0}}}$, giving the linewidth, displays a leading $T^2$ dependence, consistent with the simulation results (see inset of Fig.~\ref{fig.gap}).

\textit{Effective description.---} 
We now present an effective description capturing the key aspects of the PG mode in a significantly simpler language and with broader applicability, adapting an approach previously formulated for order-by-quantum-disorder (ObQD)~\cite{Rau2018}. 
We consider small uniform deviations (i.e., at $\vec{k} = \vec{0}$) from a classical ground state (say $\phi = 0$) with $\vec{S}_{\vec{r}}~\approx~(\sqrt{1 - \phi^2-\theta^2}, \phi, \theta)$, accurate to quadratic order in $\phi$ and $\theta$, where $\phi$ is the soft mode and $\theta$ its conjugate momentum.
For small $\phi$ and $\theta$, $\phi \approx \frac{1}{N}\sum_{\vec{r}} S^y_{\vec{r}}$ and $\theta \approx \frac{1}{N}\sum_{\vec{r}} S^z_{\vec{r}}$, with Poisson bracket $\{\phi,\theta\} = 1/N$, obtained from the canonical relation $\{S^y_{\vec{r}},S^z_{\vec{r'}}\} = \delta_{\vec{r},\vec{r'}}S^x_{\vec{r}} \approx \delta_{\vec{r},\vec{r'}}$ (taking $S^x_{\vec{r}} \approx 1$ as $\vhat{x}$ is the ordered moment direction). For this configuration, we define an effective free energy $F_{\rm eff}(\theta,\phi) = E_{\rm cl}(\theta) - T S(\phi)$, where $E_{\rm cl}(\theta)$ is the classical energy cost of nonzero $\theta$ and $S(\phi) =- \sum_{\vec{k}}\ln\omega_{\vec{k}}(\phi)$ is the entropy. 
For small $\theta$ and $\phi$, $F_{\rm eff}$ can be expanded as
 $
F_{\rm eff} \approx \frac{1}{2}N \left( \curv_{\theta} \theta^2 +  \curv_{\phi} \phi^2\right),
$
where $\curv_\theta = ({\partial^2 F_{\rm eff}}/{\partial \theta^2})/N = 2K$ and $\curv_\phi = ({\partial^2 F_{\rm eff}}/{\partial \phi^2})/N$. 
Taking $F_{\rm eff}$ as an effective Hamiltonian, the equations of motion~\cite{Goldstein2002} for $\theta$ and $\phi$ are
\begin{align}
  \frac{\del \phi}{\del t} &= +\frac{1}{N}\frac{\del F_{\rm eff}}{\del \theta} =  +\curv_{\theta} \theta, &
  \frac{\del \theta}{\del t} &= -\frac{1}{N}\frac{\del F_{\rm eff}}{\del \phi} =-  \curv_{\phi} \phi 
  \label{eq.SHM},
\end{align}
describing a harmonic oscillator. We identify the PG gap as its frequency,
$
\Delta = \sqrt{\curv_{\theta} \curv_{\phi}} 
$.
Remarkably, the $\sqrt{T}$ dependence of the PG gap is recovered, since $\curv_{\phi}$ is $O(T)$ and $\curv_{\theta}$ is $O(1)$. The curvature $\curv_{\phi}$ can be calculated from $- T S(\phi)$ within LSWT, yielding a frequency $2.46147\sqrt{T}$ for $K=5$ -- exactly the PG gap found in SCMFT as $T\rightarrow 0$ and in agreement with the spin-dynamics simulations (see Fig.~\ref{fig.gap}). 

While formulated for the Heisenberg-compass model, this line of argument can be deployed to obtain the PG gap for \emph{any} spin model exhibiting ObTD. A proof of this statement, following the strategy of Ref.~\cite{Rau2018}, will be reported elsewhere~\cite{FUTUREWORK}. %For type-I PG modes ($\cramped{\omega \propto |\vec{k}|}$, as in the Heisenberg-compass model) $\cramped{\Delta\propto\sqrt{T}}$, while for type-II modes ($\omega \propto |\vec{k}|^2$), both ${\color{blue}C_{\theta}}$, $C_{\phi}$ are $O(T)$ and thus $\Delta \propto T$. 
For type-I PG modes ($\cramped{\omega \propto |\vec{k}|}$, as in the Heisenberg-compass model we consider here), $C_{\theta}$ is $T$-independent and $C_{\phi}\sim O(T)$, giving $\cramped{\Delta\propto\sqrt{T}}$. For type-II PG modes ($\omega \propto |\vec{k}|^2$, e.g. in the ferromagnetic Heisenberg-compass model on the \emph{cubic} lattice~\cite{Rau2018}), both $\theta$ and $\phi$ are soft modes with both thus giving entropic contributions to the free energy, leading to $C_{\theta}\sim O(T)$, $C_{\phi} \sim O(T)$ and thus $\Delta \propto T$.

\begin{figure}
\includegraphics[width=\columnwidth]{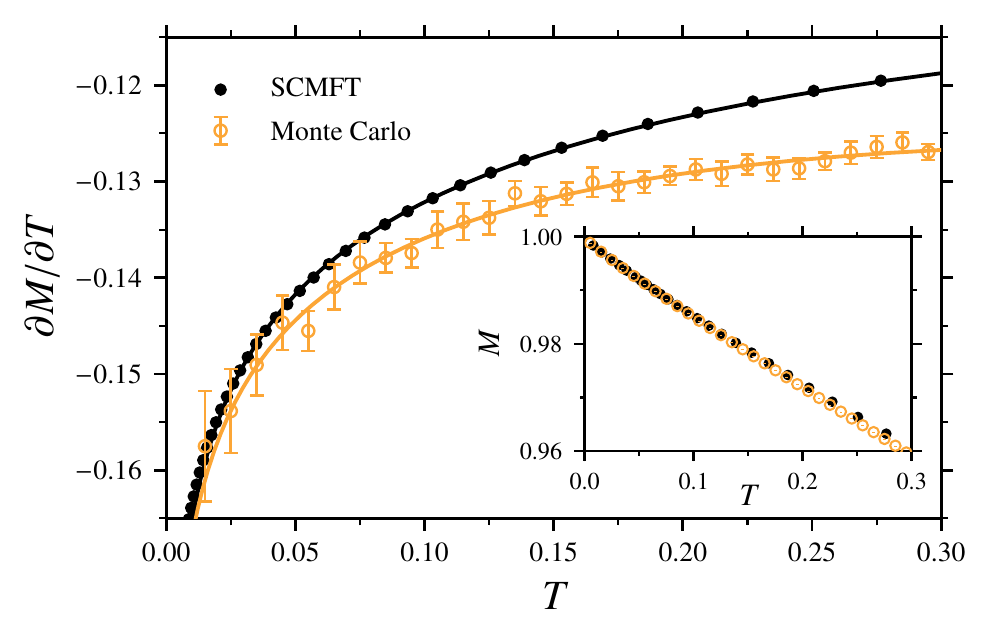}
\caption{Derivative of magnetization with respect to temperature, $\partial M/\partial T$, as a function of temperature for $L = 60$, $K =5$ using MC simulation and SCMFT. MC data is well-described by a fit motivated by SCMFT~\cite{SM}, $- 0.09815 - 0.03563 T + 0.01485\ln T$. A similar fit to SCMFT data yields $- 0.09631 - 0.01494 T + 0.01491 \ln T$. 
The inset shows $M$ as a function of temperature for the same parameters. MC error bars on $M$ are smaller than the symbol size.}
\label{fig.dMdT}
\end{figure}
\textcolor{white}{.}\\

\textit{Consequences of quasi-infrared-divergent fluctuations. ---} The ability to obtain the PG gap from LSWT presents a puzzle: the perturbative corrections $\delta A_{\vec{0}}$ and $\delta B_{\vec{0}}$ diverge logarithmically with system size~\cite{Coleman1973}, just as in the Mermin-Wagner-Hohenberg theorem~\cite{Mermin1966,Hohenberg1967}. 
How then do the curvatures of $F_{\rm eff}$ avoid these singularities and give the correct scaling? 
An analysis of the infrared divergences~\cite{SM} shows that while $\delta A_{\vec{0}}$ and $\delta B_{\vec{0}}$ are each singular, $\delta A_{\vec{0}} + \delta B_{\vec{0}}$, which determines the leading contribution to the PG gap, is \emph{finite} and reproduces the result from Eq.~(\ref{eq.SHM}). However, divergences in higher order terms do not cancel, and must be cured self-consistently~\cite{SM}.

While the divergences are mostly benign within perturbation theory for the PG gap calculations, they appear more dramatically in other quantities~\cite{Maksimov2019}, like the magnetization,
$
M =  1- \frac{1}{N}\sum_{\vec{k}} \avg{\h{a}_{\vec{k}} a_{\vec{k}}}.
$
Here, the thermal population, $\avg{\h{a}_{\vec{k}} a_{\vec{k}}}$ diverges in LSWT, rendering SCMFT necessary to obtain meaningful results. In SCMFT, the PG gap provides an infrared cutoff $\cramped{\ell \sim 1/\Delta \propto {1/\!\!\sqrt{T}} }$, giving a logarithmic contribution to $M$ scaling as $\propto T\!\ln{T}$ as $T\rightarrow 0$~\cite{SM}. The presence of this term can be diagnosed from  $\del M/\del T$, which exhibits a logarithmic singularity as $T \rightarrow 0$ for both the MC simulations and SCMFT (see Fig.~\ref{fig.dMdT}).

%%%%%%%%%%%%%%%%%%%%%%% ------------

\textit{Outlook.---} 
Our analysis of the PG gap will provide a deeper understanding of real materials exhibiting ObD. 
The existence of PG modes has been used to diagnose ObD, for example in the compounds $\mathrm{Fe}_2\mathrm{Ca}_3(\mathrm{GeO}_4)_3$~\cite{Brueckel1988}, ${\mathrm{Sr}}_{2}{\mathrm{Cu}}_{3}{\mathrm{O}}_{4}{\mathrm{Cl}}_{2}$~\cite{Kim1999} and $\mathrm{Er}_2\mathrm{Ti}_2\mathrm{O}_{7}$~\cite{Ross2014,Petit2014,Lhotel2017}. In such materials, the ObQD gap likely dominates the ObTD-induced gap discussed in this work. However, in systems where the effect of ObQD is weak or the degrees of freedom are sufficiently classical, ObTD can resurface as the leading selection effect. 
For example, our results may shed light on the rapidly growing family of two-dimensional van der Waals (vdW) ferromagnets~\cite{vdw1,vdw2,vdw3} where the ObQD gap is expected to be small and thus the contribution to the gap caused by thermal fluctuations may be more significant. 
Additionally, while reaching the classical (ObTD) selection regime is challenging in magnetic materials (due to small spin length $S$), it may be more accessible in other platforms such as those involving lattice vibrations~\cite{Han2008,Shokef2011}, dipole-coupled nanoconfined molecular rotors~\cite{dipoles-C60,water-C60,beryl,cordierite} or artificial mesoscale magnetic crystals~\cite{Dobrovolskiy2022,Keller2018,May2019,Fischer2020}. 
Whether ObTD can be realized in such topical systems, and how to detect the temperature dependent PG gap, are open questions. Our approach provides a theoretical framework and guidance for future experimental studies in this promising area of research.

\begin{acknowledgments}
We thank Itamar Aharony, Kristian Tyn Kai Chung, Alex Hickey, Daniel Lozano-G\'{o}mez, and Darren Pereira for discussions and Sasha Chernyshev and R. Ganesh for useful  comments on an early version of the manuscript.
We acknowledge the use of computational resources provided by Digital Research Alliance of Canada. This research was funded by the NSERC of Canada~(MJPG, JGR) and the Canada Research Chair Program~(MJPG, Tier I). 
\end{acknowledgments}

\bibliography{draft}

\begin{thebibliography}{98}
\bibitem [{not()}]{notefig1}%
  \BibitemOpen
  \href@noop {} {}\bibinfo {note} {The lower intensity at the band minimum
  compared to its vicinity is a numerical artifact. Due to the finite frequency
  resolution of the spin-dynamics simulations, the intensity maximum of the PG
  mode may have fallen between frequency grid points, resulting in an
  (apparent) lower intensity at the band minimum.}\BibitemShut {Stop}%
\end{thebibliography}

%%% arxiving the supplemental material as a figure
\clearpage

\addtolength{\oddsidemargin}{-0.75in}
\addtolength{\evensidemargin}{-0.75in}
\addtolength{\topmargin}{-0.725in}

\newcommand{\addpage}[1] {
\begin{figure*}
  \includegraphics[width=8.5in,page=#1]{supp.pdf}
\end{figure*}
}

\addpage{1}
\addpage{2}
\addpage{3}
\addpage{4}
\addpage{5}
\addpage{6}
\addpage{7}
\addpage{8}

\end{document}